\documentclass[fleqn,usenatbib]{mnras}

\usepackage{hyperref}
\usepackage{amsmath}
\usepackage{graphicx}
\usepackage{color}
\usepackage{ifpdf}
\usepackage[normalem]{ulem}
\usepackage{natbib}

\usepackage[T1]{fontenc}
\usepackage{ae,aecompl}
\usepackage{amssymb}	

\newcommand\mc{{\cal M}_c{}}

\newcommand\msun{M_\odot}
\newcommand\unit[1]{\text{#1}}
\newcommand\abbrvPSgrb{PSG}
\newcommand\abbrvPSellipticals{PSE}
\newcommand\abbrvKBLowZa{BD2010}
\newcommand\ForInternalReference[1]{}

\newcommand\cqg{CQG}
\newcommand\ExcitingGalaxy{h258}
\newcommand\BoringGalaxy{h277}
\newcommand\DwarfOne{h603}
\newcommand\DwarfTwo{h516}

\title[Effects of host galaxy properties on LIGO binaries]{The effects of host galaxy properties on merging compact binaries detectable by LIGO}

\author[R. O'Shaughnessy et al.]{R. O'Shaughnessy,$^1$\thanks{E-mail: oshaughn@mail.rit.edu}
J. M. Bellovary,$^2$ 
A. Brooks,$^3$
S. Shen,$^4$ 
F. Governato,$^5$ 
C. R. Christensen$^6$ \\
$^1$Center for Computational Relativity and Gravitation, Rochester Institute of Technology, Rochester, NY 14623, USA  \\
$^2$Department of Physics, Queensborough Community College, Bayside, NY 11364, USA\\
$^3$Department of Physics and Astronomy, Rutgers University, Piscataway, New Jersey, 08854, USA\\
$^4$Institute of Astronomy, University of Cambridge, Cambridge CB3 OHA, UK\\
$^5$Department of Astronomy, University of Washington, Seattle, WA 98195, USA\\
$^6$Department of Physics, Grinnell College, Grinnell, IA, 50112, USA\\
 }

\date{Accepted XXX. Received YYY; in original form ZZZ}

\pubyear{2016}

\begin{document}
\label{firstpage}
\pagerange{\pageref{firstpage}--\pageref{lastpage}}
\maketitle

\begin{abstract}

Cosmological simulations of galaxy formation can produce present-day galaxies with a large range of assembly and star formation histories.  A detailed study of the metallicity evolution and star formation history of such simulations can assist in predicting LIGO-detectable compact object binary mergers.
Recent simulations of compact binary evolution suggest the compact object merger rate depends
sensitively on the progenitor's metallicity.  Rare low-metallicity star formation during galaxy
assembly can produce more detected compact binaries than typical star formation.   
Using detailed simulations of galaxy and chemical evolution, we determine how sensitively the compact binary populations
of galaxies   with similar
present-day appearance depend on the details of their assembly.
We also demonstrate by concrete example the extent to which dwarf galaxies overabundantly produce compact binary
mergers, particularly binary black holes, relative to more massive galaxies. 
We discuss the implications for transient multimessenger astronomy with compact binary sources.

\end{abstract}




\section{Introduction}

Gravitational waves have been detected from  coalescing  black hole binaries \citep{DiscoveryPaper,LIGO-O1-BBH}. 
Over the next few years, ground-based gravitational wave detectors like LIGO 
and Virgo  should detect the gravitational
wave signal from many more similar merging compact binaries \citep{LIGO-O1-BBH,RatesPaper,gwastro-EventPopsynPaper-2016}, as well as  binary  neutron stars and black hole-neutron star binaries \citep{LIGO-Inspiral-Rates,popsyn-LowMetallicityImpact2c-StarTrackRevised-2014}.
The host galaxies of gravitational wave sources will be identified, either directly or statistically.  
If the event is a merger of at least one neutron star, it is expected to be accompanied by detectable
electromagnetic radiation  via a number of  mechanisms \citep[see][and references therein]{Metzger16} in addition to the strong gravitational wave signal. 
A multimessenger detection will pin down
the sky position and therefore approximate birthplace of each merging binary \citep{Nissanke13}.    
Even in the absence of well-identified electromagnetic counterparts, host galaxy information is still available from gravitational wave localization alone \citep{2016LRR....19....1A,2016arXiv160307333S}.  As GW detector
networks increase in number and sensitivity, these localizations will allow statistical and, eventually, unique identification of
host galaxies directly, even without associated electromagnetic emission.
As with supernovae and  GRBs, these host galaxy associations   are expected to tightly constrain models for compact binary
formation; see, for comparison,  \citet{2011MNRAS.412.1508M}, \citet{long-grb-GuettaPiran2007},
\citet{2014ARAA..52...43B}, and references therein. 

The host galaxies of distant short GRBs have already been extensively investigated, with the associations being used to
draw preliminary conclusions about their progenitors  \citep{2014ARAA..52...43B}.   
Unlike GRBs, detected gravitational wave sources will be limited by the range of LIGO to the local universe; for
example,  binary neutron star sources should be closer than $400\unit{Mpc}$.  Due to their proximity, each host galaxy
can be explored at great depth and detail via  position-resolved spectroscopy, enabling detailed position-resolved star-formation
and chemical evolution histories  \citep[see,e.g.][]{2009MNRAS.396..462K,2014MNRAS.444..336C,
CALIFA,CALIFA2}.  
However, unlike short GRBs and supernovae (SN), present-day compact  binary populations can depend sensitively on rare low metallicity
star formation.
In this work, we assess by concrete example the extent to which detailed analysis of individual galaxies' assembly histories will be essential in
investigating key physical questions about the origin of compact binary mergers.

This paper is organized as follows. 
In \S \ref{sec:sims}, we describe detailed hydrodynamical simulations of  several galaxies, including four of Milky Way-mass and two dwarfs.  Though the four Milky Way-like galaxies are morphologically similar at $z = 0$, their star formation and chemical evolution history have subtle differences due to their
distinctive merger histories.   To demonstrate the practical impact of
these differences as well as that of halo mass,  in \S \ref{sec:model}, we introduce a simple,
metallicity-dependent  phenomenological model to calculate the present-day rate and mass distribution of compact binary
mergers from  a galaxy's known history.   In \S \ref{sec:results:BBH}, we use this model to investigate  the compact
binary  coalescence rate
dependence on each galaxy's assembly history and mass.   
In \S~\ref{sec:Discussion} we discuss the implications of our study for the interpretation of host galaxy associations
identified via transient multmiessenger astronomy, in the short and long term.  We summarize our results in \S~ \ref{sec:conclude}.

\section{Cosmological simulations}
\label{sec:sims}

\subsection{Simulating galaxy evolution}
To thoroughly study the significance of low metallicity star
formation, we examine cosmological smoothed particle hydrodynamics
(SPH) $N$-body simulations of Milky Way-like galaxies with GASOLINE
\citep{Stadel01,Wadsley04}.  These simulations allow us to analyze both
spatially and temporally resolved star formation, and determine the metallicity history of compact object progenitors.

We selected our simulated regions of interest from a volume of
uniform resolution, and resampled the region at very high
resolution using the volume renormalization technique \citep{Katz93}.
This technique allows us to follow the detailed physical processes
involved in galaxy evolution in our selected region while still
including large-scale torques from cosmic structure.  Our cosmological parameters are $\Omega_0 = 0.24$, $\Omega_{\rm baryon} = 0.04$, $\Lambda = 0.76$, h = 0.73, $\sigma_8 = 0.77$ 
 \citep{WMAP3}.  \footnote{Since we are simulating individual galaxy environments rather than large populations of halos, the selection of cosmological parameters provides a negligible contribution to the variance in the overall evolutionary history of galaxies in our simulations. } We model the ionizing UV background
with the prescription from \citet{Haardt96}.  Our interstellar medium (ISM) model includes the non-equilibrium formation and destruction of H$_2$, which is incorporated in the cooling model along with metal lines, along with shielding of HI and H$_2$ \citep{Christensen12}.  Stars form
probabilistically from gas particles which meet density ($n_{min} =
0.1$ amu cm$^{-3}$) and temperature ($T_{max} = 10^3$ K) thresholds, though since star formation also depends on the H$_{\rm 2}$ content of a particle (see below) the densities are nearly always much higher than this threshold. If a gas
particle meets these criteria, it has a likelihood of forming a star
particle (representing a simple stellar population with a Kroupa IMF
\citep{Kroupa93}) which is given by

\begin{equation}
p = \frac{m_{\rm gas}}{m_{\rm star}} (1 - e^{c^*\rm X_{\rm H_2} \Delta t/t_{\rm form}})
\end{equation}

\noindent
where the star formation efficiency parameter $c^*$ is set to 0.1 such
that our galaxies match the observed Kennicutt-Schmidt law
\citep{Kennicutt89};  X$_{\rm H_2}$ is the molecular hydrogen fraction of the gas particle; $m_{star}$ and $m_{gas}$ are the star and gas
particle masses; \footnote{Gas particles start with a set mass
 and may gain mass from feedback and lose it to star formation.  Each star particle, when
  formed, has $1/3$ of the progenitor mass of the forming gas particle.  See \cite{Christensen10} for a discussion
of resolution issues in SPH simulations.} 
 $t_{form}$ is the dynamical time for the gas
particle; and $\Delta t$ is the time between star formation episodes,
which we set to 1 Myr.  A detailed study of different ISM models and the resulting star formation properties by \citet{Christensen14a} demonstrates that this model allows star formation to occur in clumps of dense gas, comparable to giant molecular clouds.  

We model supernova feedback using the
blastwave formalism described in \citet{McKee77} and implemented in
our simulations as in \citet{Stinson06}.  Each supernova releases
$E_{SN} = 10^{51}$ erg into the surrounding gas with a radius
determined by the blastwave equations.  These particles are not
allowed to cool for the duration of the blastwave, mimicking the
adiabatic expansion phase of a supernova explosion.  Previous works have found
that this set of parameters results in realistic galaxies which obey a
number of observed relations such as the mass-metallicity relation
\citep{Brooks07,Christensen16}, 
 the Tully-Fisher relation \citep{Christensen16}, and
the size-luminosity relation \citep{Brooks11}, as well as reproduce
the detailed characteristics of bulgeless dwarf galaxies
\citep{Governato10,Governato12}, low-mass disk galaxies with bulges \citep{Christensen14b}, and the Milky Way \citep{Guedes11}.

Metals are created in supernova explosions
and deposited directly to the gas within the blast radius.  Stellar
masses are converted to ages as described by \citet{Raiteri96}, and
stars more massive than 8 M$_\odot$ are able to undergo a Type II
supernova.    For Type II supernovae, iron and oxygen are produced according to the analytic fits used in \citet{Raiteri96} using the yields from \citet{Woosley95}:

\begin{equation}
M_{\rm Fe} = 2.802 \times 10^{-4} M_*^{1.864}
\end{equation}
\begin{equation}
M_{\rm O} = 4.586 \times 10^{-4} M_*^{2.721}.
\end{equation}

Feedback from Type Ia supernovae also follows \citet{Raiteri96}.  Each supernova produces $0.63\msun$ iron and $0.13 \msun$ oxygen \citet{Thielemann86}.  Metal production from stellar winds is also included; we implement stellar wind feedback based on \citet{Kennicutt94} and the returned mass fraction is derived using a function by \citet{Weidemann87}.  The returned gas has the same metallicity as the star particle.

Also included in our simulations is a scheme for turbulent metal
diffusion \citep{Shen10}.   Once created,  metals
diffuse through the surrounding gas, according to 
\begin{equation}
\frac{dZ}{dt}|_{diff} = \nabla (D \nabla {Z}) \\
\end{equation}
where the diffusion parameter $D$ is given by
\begin{equation}
D = C_{diff} |S_{ij}| h
\end{equation}
and $h$ is the SPH smoothing length, $S_{ij}$ is the trace-free shear
tensor, and $C_{diff}$ is a dimensionless constant which we set to a conservative value of 0.03.  
Combined with infall, this procedure produces a range of metallicities within each galaxy.

We do not include specific prescriptions for metal distribution based on initial metallicity (such as for Population III stars) or variations in IMF.  We cannot thus discuss Population III contributions to the gravitational wave background; however, studies by  \citet{Hartwig16}  and \citet{Dvorkin16} suggest that this contribution is fairly negligible.  We have shown that our scheme for metal production and distribution produces galaxies which match the mass-metallicity relation at $z \sim 3$ \citep{Brooks07} as well as in the local universe \citep{Christensen16}.  At higher redshifts, there is a very large spread in the metal distributions of damped Lyman alpha systems \citep{Dvorkin15}; manifestly, other features in the galaxy evolutionary history play a role in its present day metallicity.  Recently \citet{Hunt16a,Hunt16b} has demonstrated a relationship between of the mass, metallicity, and star formation rate, which demonstrates an evolutionary relationship between these quantities.  As we will discuss in future work, we have verified that our evolutionary prescriptions are also qualitatively consistent with this relation.

We identify individual galaxies using the halo finder $AHF$
\citep{Gill04,Knollmann09}, which identifies haloes based on an
overdensity criterion for a flat universe \citep{Gross97}.  In each simulation, we are focusing on the stars which make up the primary (i.e. most massive) galaxy within the zoomed-in high resolution region at $z = 0$.

\subsection{Milky-Way-like galaxies with distinct histories }

\begin{figure*}
\includegraphics[width=\columnwidth]{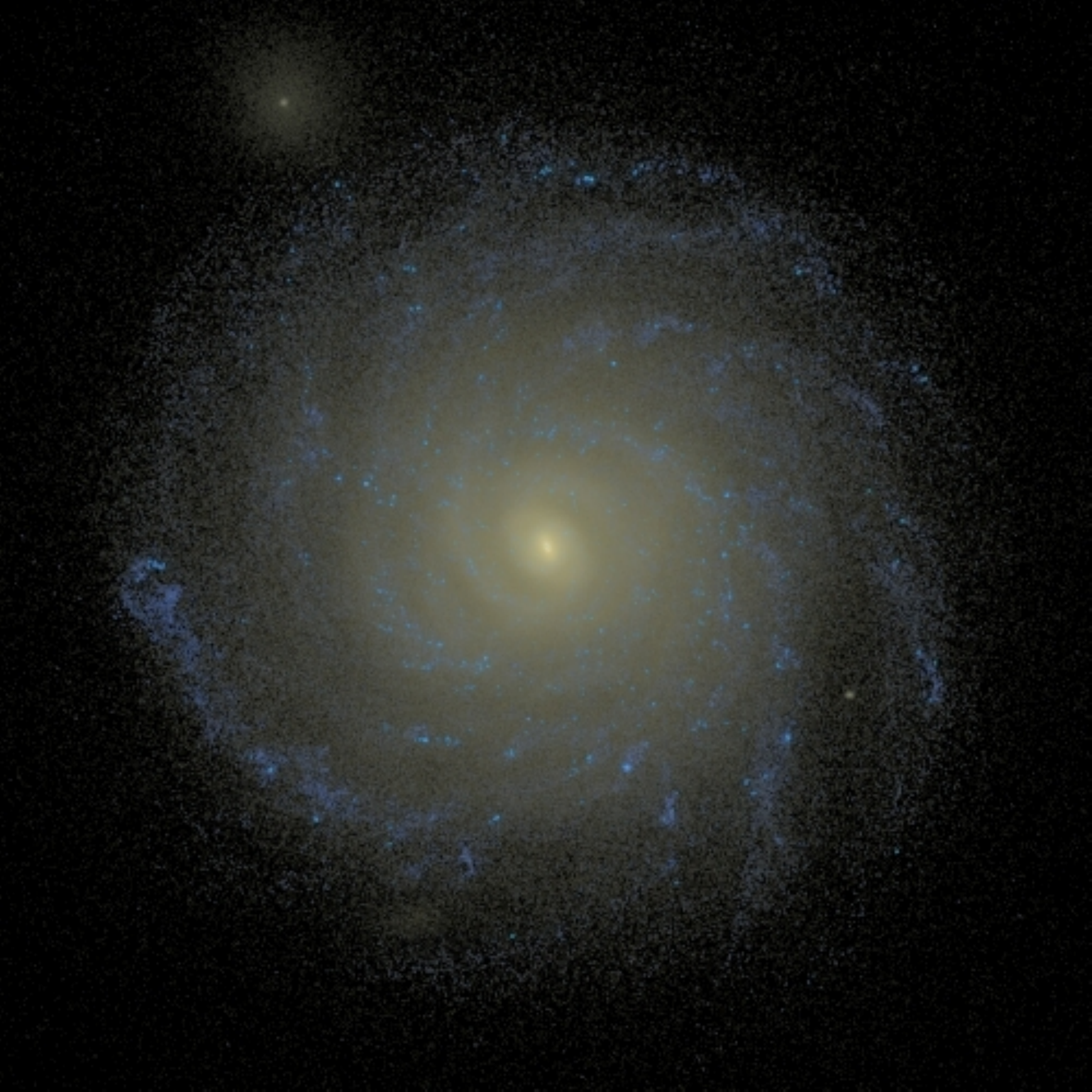}
\includegraphics[width=\columnwidth]{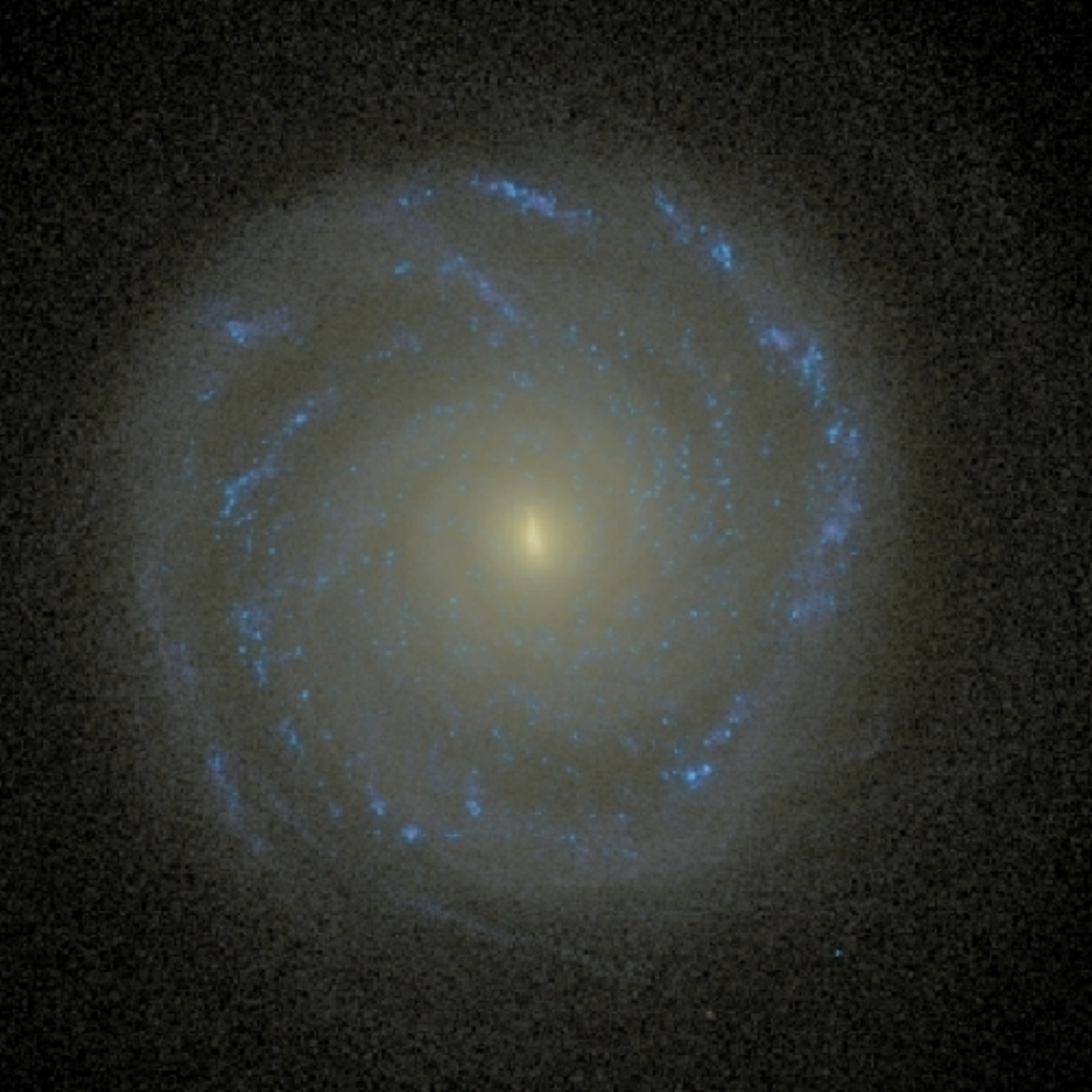}
\caption{\label{fig:GalaxyImages}Synthetic SDSS $gri$ images of two of our Milky Way-type galaxies, h277 (left) and h258 (right), created with \textsc{SUNRISE} \citep{Jonsson06}.
\label{fig:images}
}
\end{figure*}

The evolution
of a galaxy, in terms of its stellar mass and metallicity evolution,
depends strongly on its interaction history.  Galaxies which appear
similar at the present day may have had drastically different
histories, which may result in differences in compact object merger
rates.  To investigate whether galaxy history affects the compact binary event
rate, we have chosen four simulations which are morphologically similar at $z
= 0$  (see Figure \ref{fig:images}) but differ strongly in their merger histories.

The simulation h277  is a Milky Way analog with a
quiescent merger history.  It experiences its last major merger at $z
= 3$, after which a small number of minor interactions permeate its
life.  This simulation has been shown to emulate several Milky Way
properties, including stellar dynamics
\citep{Loebman12,Loebman14,Kassin14}, baryon fraction
\citep{Munshi13}, and satellite properties \citep{Zolotov12,Brooks14}.  It has a virial mass of $M_{vir} =
6.79 \times 10^{11} \msun$, stellar mass $M_* = 4.24 \times 10^{10}
\msun$, and maximum circular velocity $v_{\rm circ} = 235$ km s$^{-1}$.

  The simulation h258, on the other hand, has a much
  more active merger history.  At $z = 1$ there is a 1:1 merger event;
  a gas disc rapidly reforms following the collision (see
  \citet{Governato09}), resulting in a massive galaxy at $z = 0$
  which looks remarkably similar to the Milky Way and to the other
  simulation, h277.  Prior to the $z = 1$ merger, each of the four
  progenitor galaxies actually experiences its own additional major
  merger events around $z = 3$.  The combination of
  the series of major mergers, plus a number of minor interactions and
  flybys, gives a stark contrast to the relatively quiescent history
  of h277.  At $z = 0$, h258 has a virial mass of $M_{vir} = 7.74
  \times 10^{11} \msun$, stellar mass $M_* = 4.46 \times 10^{10}
  \msun$, and maximum circular velocity $v_{\rm circ} = 242$ km s$^{-1}$.  
  
We include two additional Milky Way simulations with similar $z = 0$ properties, which have evolutionary histories that fall in between the extremes of the two described above.  The galaxy h239 has a total virial mass of $M_{vir} = 9.3 \times 10^{11}\msun$, stellar mass $M_* = 4.50 \times 10^{10}
\msun$, and maximum circular velocity $v_{\rm circ} = 250$ km s$^{-1}$.  The galaxy h285 has a total virial mass of $M_{vir} = 8.82 \times 10^{11}\msun$, stellar mass $M_* = 4.56 \times 10^{10} \msun$, and maximum circular velocity $v_{\rm circ} = 248$ km s$^{-1}$.  These galaxies are also described in \citet{Bellovary14,Sloane16}.  

\begin{figure*}
\includegraphics[width=\columnwidth]{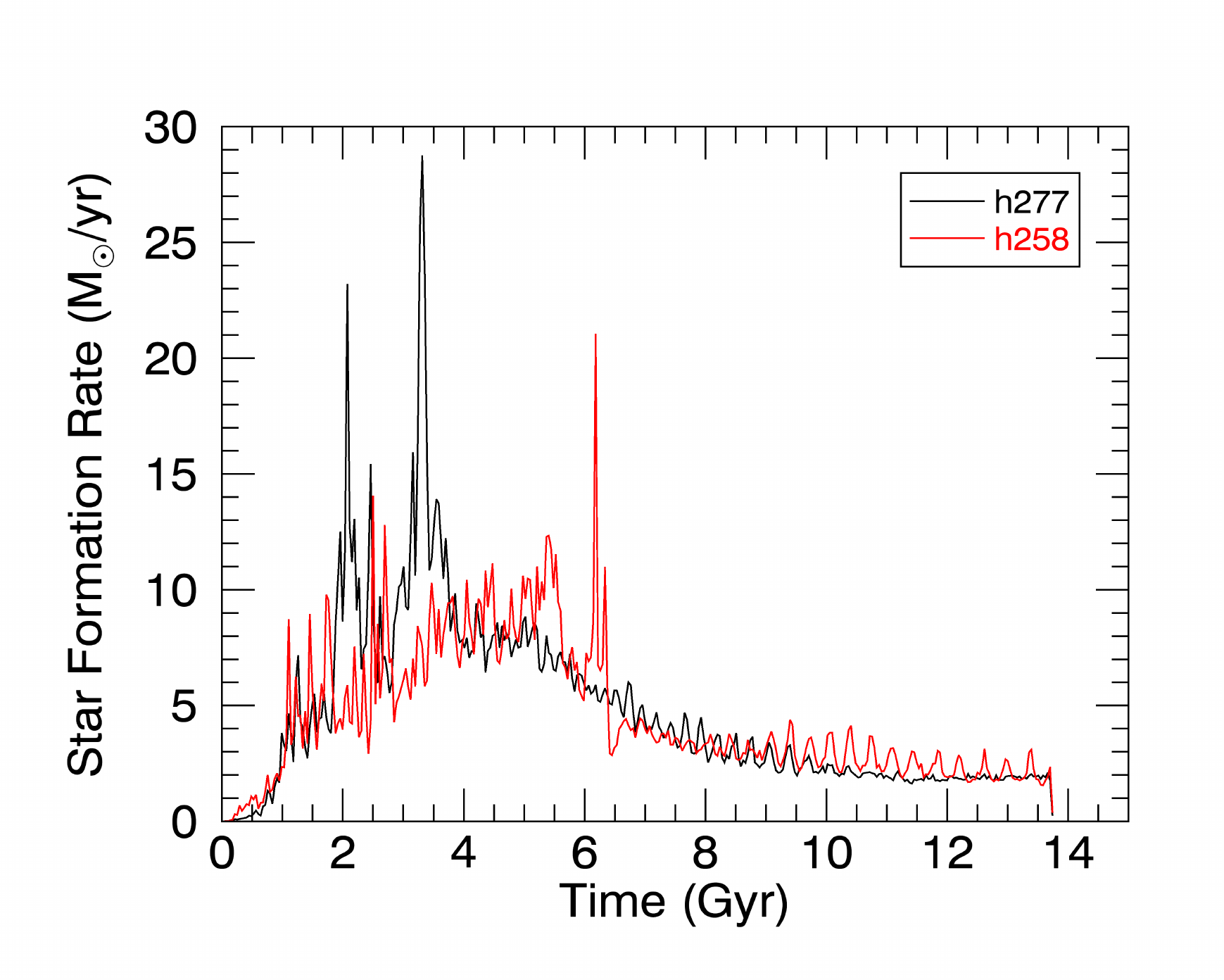}
\includegraphics[width=\columnwidth]{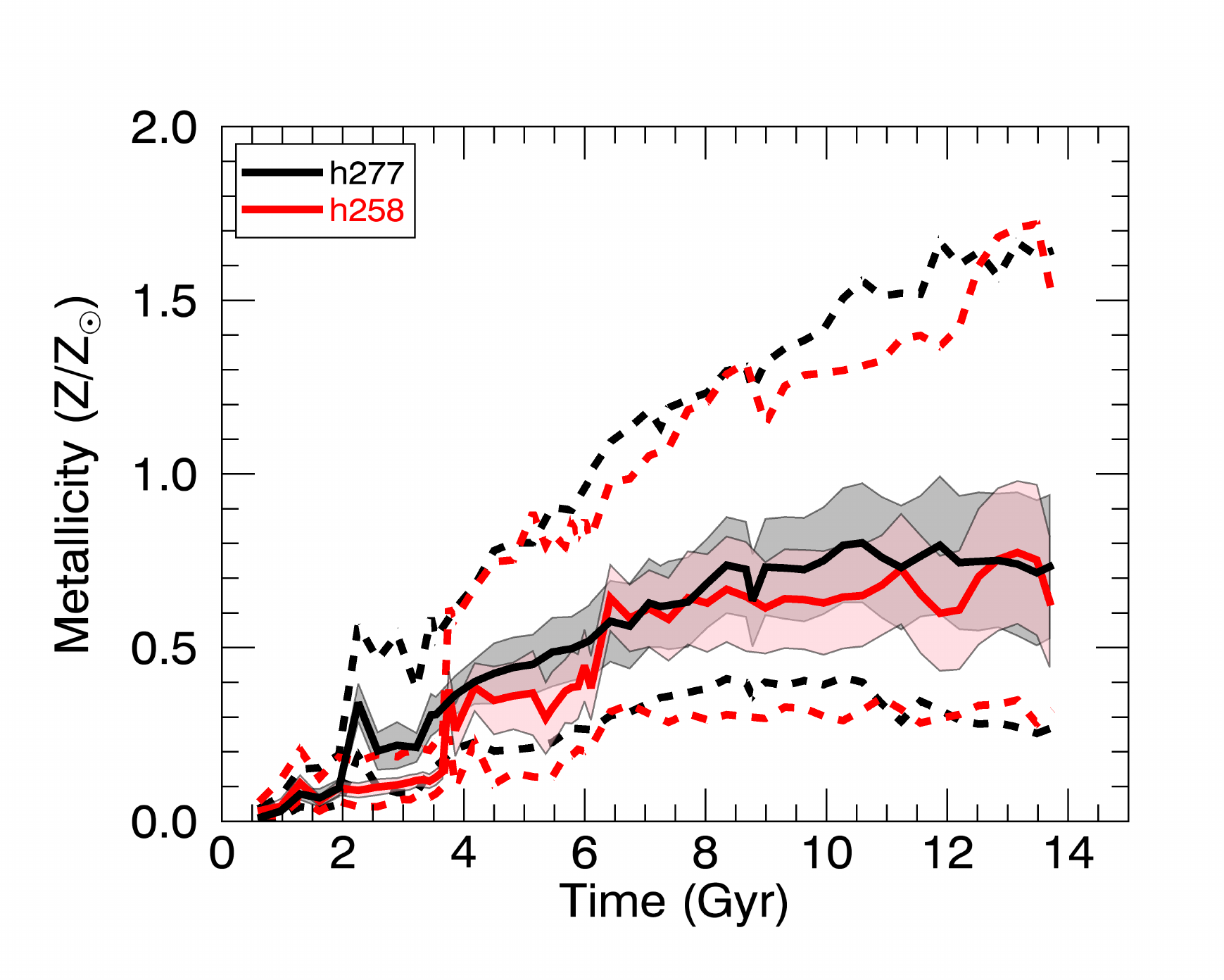}
\caption{\label{fig:TwoGalaxies}\textbf{Star formation and metallicity versus time}:  \emph{Left panel}: Star formation history $\dot{M}_{*}$
  versus time.  Black corresponds to \BoringGalaxy{} and red to \ExcitingGalaxy.  
\emph{Right panel}: A plot of the  metallicity $Z$ of recently-formed stars versus time.  Solid red and black lines show
the mean metallicity; dotted lines correspond to  90\% of newly-born stars with lower metallicity,   or 10\% of
newly-born stars with greater metallicity.  The shaded region represents one standard deviation.  When these two galaxies are $2-3\unit{Gyr}$ old, prior to the first major
merger, newly-born stars are created with significantly different metallicity.  Additionally, prior to the
second major merger of h258 at $\simeq 6\unit{Gyr}$, stars form in the \ExcitingGalaxy{} galaxy at a systematically lower
metallicity than in the \BoringGalaxy{} counterpart.  Even from 8-13 Gyr, h258 has a slightly lower metallicity than h277.
}
\end{figure*}

  Due to the differences in merger histories, the star formation histories and metallicity
  evolution of h258 and h277 also differ at early times.  Figure
  \ref{fig:TwoGalaxies} shows the star formation history (left panel) and metallicity evolution (right panel) of h277 (black) and h258 (red).\footnote{We choose not to show galaxies h239 and h285 in our figures, as they fall in between the values bracketed by h258 and h277 and add confusion to the plots.}  The star formation histories are quite different at early times, where h277 has larger bursts of star formation between 2-4 Gyr, but h258 has a large burst at $\sim 6$ Gyr during a major merger.   The right panel shows the mean metallicity of
  recently formed stars vs time (thick solid lines), where we define
  recent as within the past 50 Myr.  The shaded regions correspond to
  one standard deviation of the mean, while the thick dashed lines
  represent the top and bottom 90\%.  We see that the early evolution
  of the metal properties of these galaxies does differ - between 0.5
  and 6 Gyr, h277 hosts a \emph{modestly}  more metal-rich population than
  h258.   For most astrophysical processes, the small metallicity difference illustrated here will have no impact on
  present-day observables.  However,  the population of binary black holes  depends sensitively on all low metallicity star formation
  over cosmic time.

We wish to point out that these simulations do not include the effects of supermassive black holes (SMBHs).  These galaxies are at the mass where feedback from SMBH accretion is thought to affect star formation, and adding these effects may add additional scatter to the stellar and metal evolution of each galaxy, which could alter the binary black hole merger history as well.

\subsection{Dwarf galaxies}

We have also employed the results of detailed simulations for two dwarf galaxies: \DwarfOne{} and \DwarfTwo{}.  The simulation of h603 consists of a low-mass disc galaxy (qualitatively similar to M33).  It has a virial mass of $3.4 \times 10^{11} \msun$, stellar mass of $7.8 \times 10^9 \msun$, and maximum circular velocity of 111 km s$^{-1}$.  The structure and star formation of this galaxy has been extensively studied by \citet{Christensen14b}.  We also include a bulgeless dwarf galaxy, h516, which has a disc with irregularly distributed star formation, with a virial mass of $3.8 \times 10^{10}\msun$, stellar mass $2.5 \times 10^8 \msun$, and maximum circular velocity of 65 km s$^{-1}$.  Images of these galaxies are shown in Figure \ref{fig:dwarfimages}, and we show their star formation history and metallicity evolution in  Figure \ref{fig:TwoDwarfGalaxies}.  Note that these galaxies have quite different masses, and are not meant to be directly comparable.  The more massive h603 has a much more active star formation history and an overall increasing metallicity with time, whereas the less massive h516 is characterized by small bursts of star formation and a fairly flat metallicity evolution, perhaps due to its substantial outflows.

\begin{figure*}
\includegraphics[width=\columnwidth]{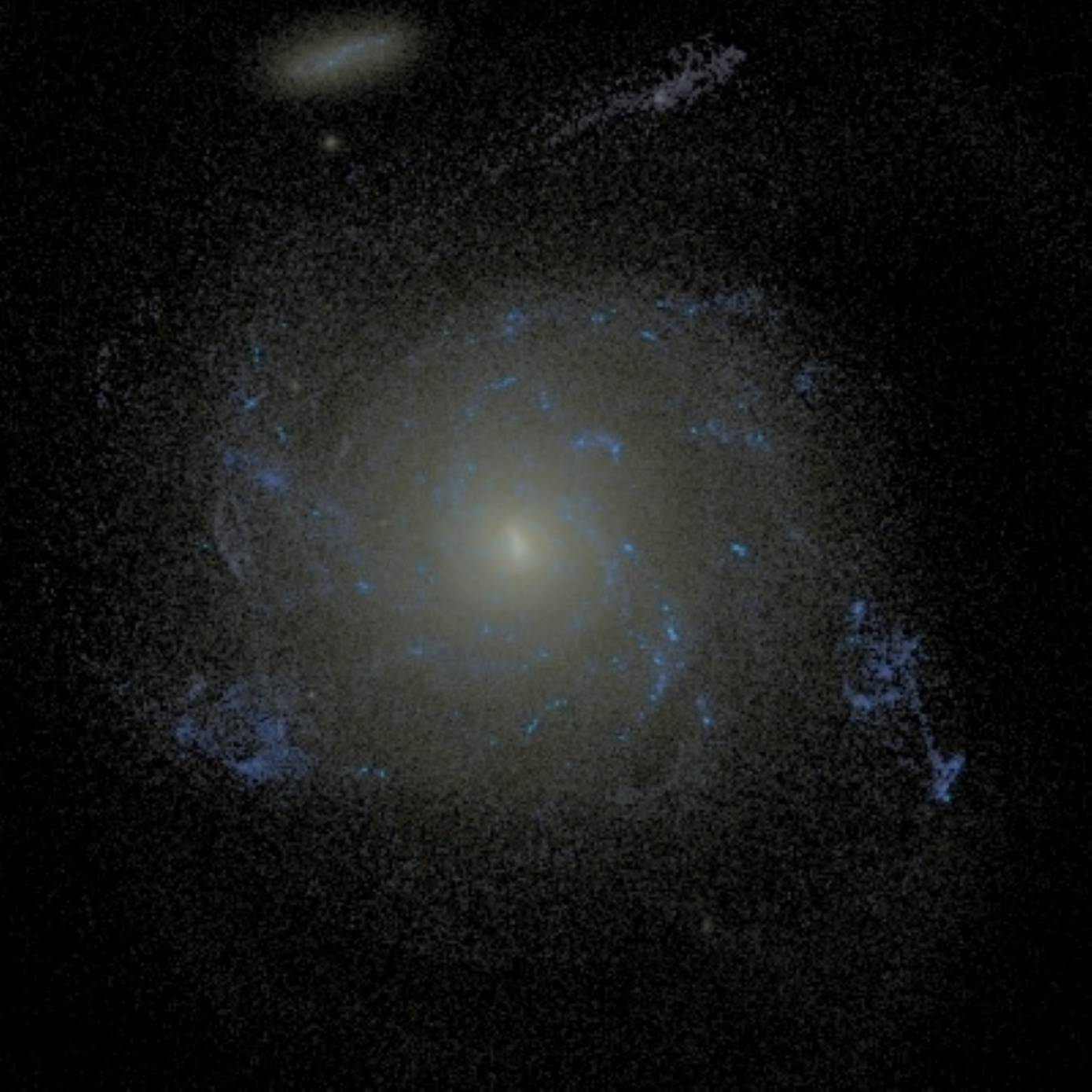}
\includegraphics[width=\columnwidth]{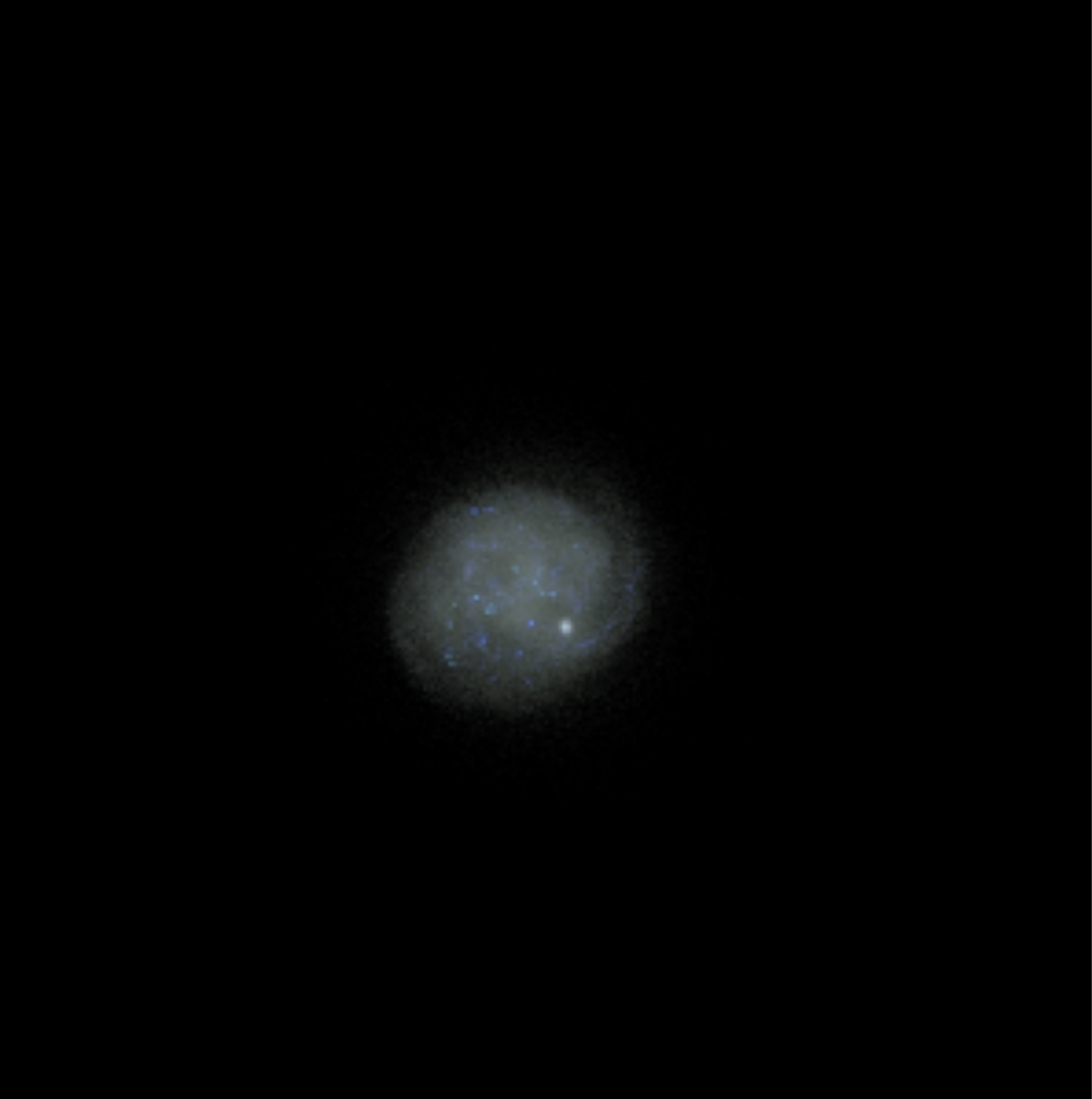}
\caption{\label{fig:dwarfimages}Synthetic SDSS $gri$ images of our two lower-mass galaxies, h603 (left) and h516, created with \textsc{SUNRISE} \citep{Jonsson06}.
\label{fig:dwarfimages}
}
\end{figure*}

\begin{figure*}
\includegraphics[width=\columnwidth]{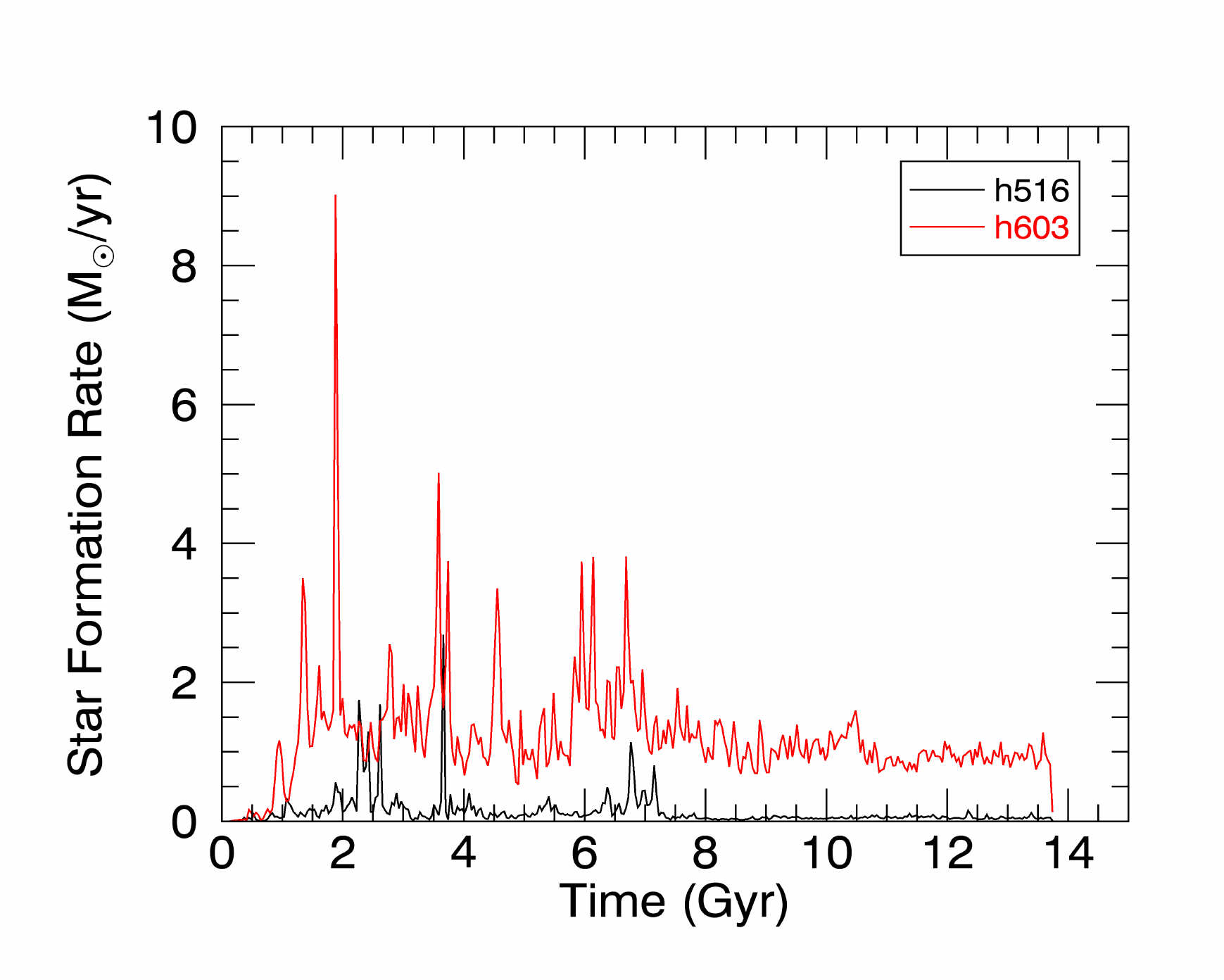}
\includegraphics[width=\columnwidth]{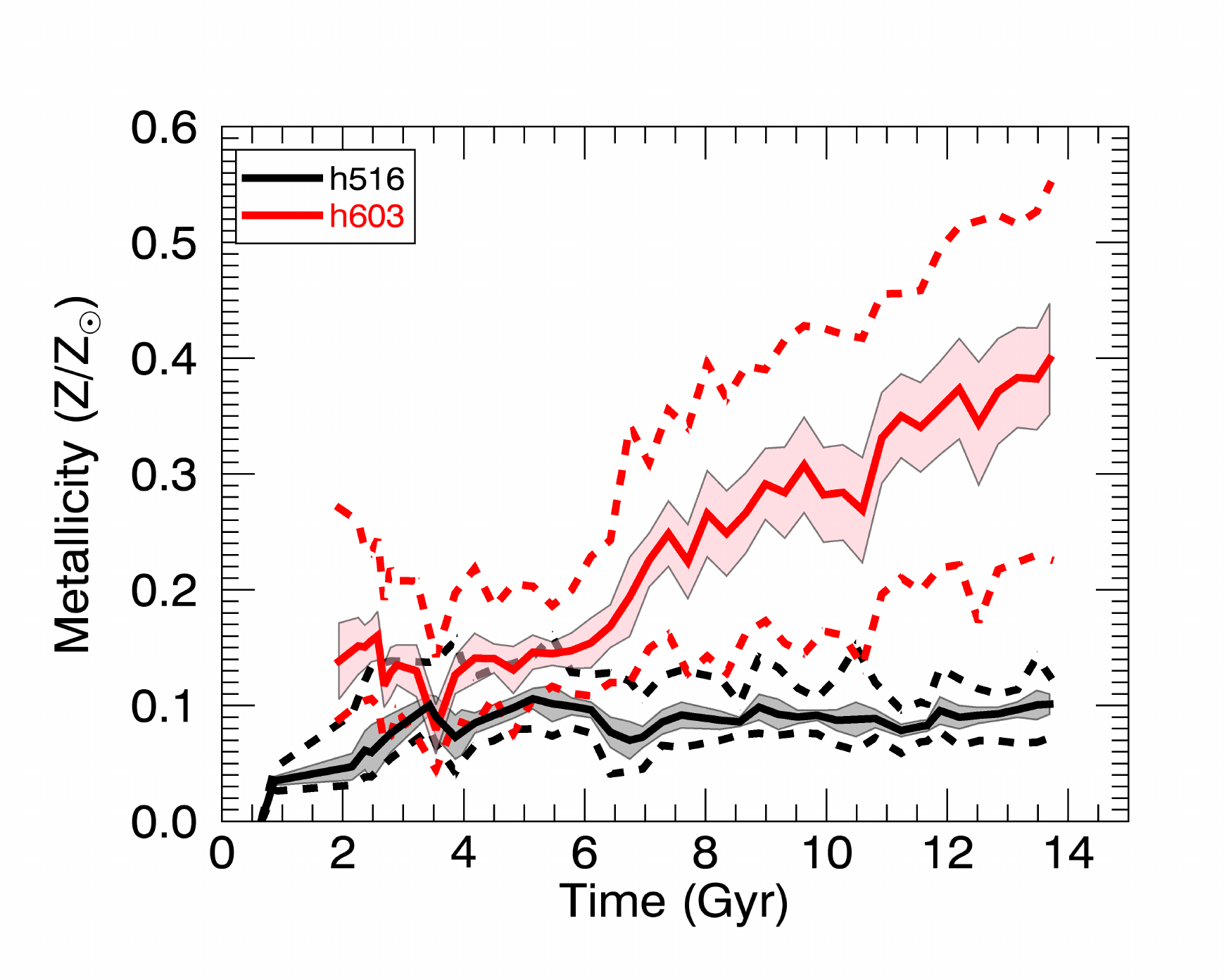}
\caption{\label{fig:TwoDwarfGalaxies}\textbf{Star formation and metallicity versus time}:  \emph{Left panel}: Star formation history $\dot{M}_{*}$
  versus time.  Red corresponds to h603 and black to h516.  
\emph{Right panel}: A plot of the  metallicity $Z$ of recently-formed stars versus time.  Solid red and black lines show
the mean metallicity; dotted lines correspond to  90\% of newly-born stars with lower metallicity,   or 10\% of
newly-born stars with greater metallicity.  The shaded region represents one standard deviation.
}
\end{figure*}

\section{Detection-weighted compact binary formation}
\label{sec:model}

Our goal in this work is to estimate the \emph{ratio} of compact binaries that should be merging, at present, in the four
simulated Milky-Way analog galaxies described above as well as the dwarfs.

To explore plausible binary detection scenarios, we adopt a parametrized formalism for binary evolution and event
detection in an individual galaxy, motivated by
the detailed studies of  \cite[][(hereafter \abbrvPSgrb{})]{PSgrbs-popsyn} and  \citet[][(hereafter \abbrvPSellipticals{})]{PSellipticals};  similar approaches have been used by
\cite{2016arXiv160508783L} and others.  Since prior population synthesis investigations  extend relatively smoothly to very metallicity, we do not introduce a
distinct, special group of low-metallicity Population III stars.

Binary evolution calculations suggest the binary compact object formation rate depends sensitively on the assumed metallicity, in conjunction
with other parameters \citep[see,\,e.g.][and references
  therein]{popsyn-LowMetallicityImpact-Chris2008,popsyn-LIGO-SFR-2008,gwastro-EventPopsynPaper-2016}.
Gravitational wave detectors are also far more sensitive to  massive compact binaries, which are  preferentially  formed in low metallicity
environments \citep{PSellipticals,popsyn-LowMetallicityImpact2c-StarTrackRevised-2014}.  As a result, low metallicity
environments can be overwhelmingly efficient factories for detectable black hole binaries
\citep{popsyn-LowMetallicityImpact2c-StarTrackRevised-2014,gwastro-EventPopsynPaper-2016}.  
For this reason, our estimates for compact binary coalescence rates must account for how often  different star-forming
conditions occur;  how often compact binaries that can coalesce now can derive from each environment; and how often LIGO will detect compact
binaries with different masses, all other things being equal.

To characterize how much more likely LIGO will detect coalescing binaries with different masses, we use a common and
naive estimate for the volume to which advanced LIGO is sensitive \citep[see,e.g.,][]{PSellipticals}:\footnote{For
  simplicity, in this calculation we neglect the effects of cosmology; strong field coalescence; and black hole spin;
  see \cite{popsyn-LowMetallicityImpact2c-StarTrackRevised-2014}, \cite{AstroPaper}, or \cite{RatesPaper} for more details.}
\begin{eqnarray}
V = \frac{4\pi}{3} [445 \unit{Mpc}]^3 \int p(m_1,m_2|Z) dm_1 dm_2 [ (\mc/1.2 M_\odot)^{5/6}]^3
\end{eqnarray}
where $\mc\equiv (m_1 m_2)^{3/5}/(m_1+m_2)^{1/5}$.   
This expression depends explicitly on an assumed and metallicity-dependent mass distribution for compact object
mergers, through the characteristic chirp mass $\mc_{*}(Z)\equiv [\int p(m_1,m_2|Z) dm_1 d,m_2 (\mc/1.2
  M_\odot)^{15/6}]^{5/16}$ .  We calibrate our metallicity-dependent mass distributions to metallicity-dependent binary
evolution calculations presented in \cite{popsyn-LowMetallicityImpact2-StarTrackRevised-2012} and
\cite{popsyn-LowMetallicityImpact2c-StarTrackRevised-2014}.  For neutron
stars, we adopt a fiducial neutron star mass of $1.4 M_\odot$ at all metallicities.  For BH-NS binaries, we adopt a
highly simplified model: the black hole masses are uniformly  drawn from $5 M_\odot$ to the maximum 
black hole mass $M_{max}(Z)$ allowed by
\emph{isolated} stellar evolution, as reported in prior work \citep[see,e.g.][and references
  therein]{gwastro-EventPopsynPaper-2016}.  In agreement with much more detailed prior work \cite{popsyn-LowMetallicityImpact2c-StarTrackRevised-2014}, this simplified model yields typical chirp masses for BH-NS binaries that
vary slightly as $Z/Z_\odot$ decreases, from a lower limit of $3 M_\odot$ near solar metallicity to an upper limit of
$4.3 M_\odot$ in low-metallicity environments.  Not least because the average chirp mass for BH-NS simply cannot vary
dramatically, given the functional form of $\mc$, our results for the BH-NS coalescence rate per unit galaxy mass do not depend sensitively on the choice of black hole mass
distribution.  Finally, for binary black holes, we assume  comparable-mass binaries
form ($m_1=m_2$), with the distribution of component masses chosen to be $\propto m_1^{-p}$ 
 between $5 M_\odot$ and $M_{max}(Z)$ and zero otherwise, adopting a fiducial exponent $p=2$; see, e.g.,
\cite{popsyn-LowMetallicityImpact2b-StarTrackRevised-2013}.    For binary black holes, this means  the
detection-weighted mass distribution [$\propto p(m_1,m_2)
\mc^{15/6}$] depends weakly on mass [$ p(m_1,m_2)
\mc^{15/6} \propto m_1^{-0.5} \delta(m_1-m_2)$.  In our simple model, the total  and chirp mass
distributions   are qualitatively consistent with detailed  binary evolution calculations \citep{popsyn-LowMetallicityImpact2b-StarTrackRevised-2013}.

To characterize the frequency of different star-forming conditions, we use our cosmological simulations of Milky Way-like galaxies, which provide each galaxy's star formation rate
$\dot{M}_*$ and metallicity distribution $p(\log Z|t)$ over time (see Eqn. \ref{eqn:rate}).

To characterize how often compact binaries form and coalesce, we use the ansatz adopted in \abbrvPSellipticals{} and
\abbrvPSgrb{}: for a star-forming parcel of mass $\Delta M$ the number of binaries born at time $0$ which
are coalescing now is $\lambda \Delta M dP_t/dt$, where $\lambda$ is an overall efficiency per unit mass and $P_t(<t|Z)$ is
a metallicity-dependent delay time distribution.\footnote{For simplicity, we assume the mass and delay time
  distributions are uncorrelated.  Figure A9 in \abbrvPSgrb{} shows this approximation, while not strictly true, is an
  excellent approximation for merging BH-BH binaries at solar metallicity.  }
    In this paper we are investigating the relative contribution of
different star-forming environments and galaxy evolutionary histories, not the overall normalization, so the overall scale of $\lambda$ is irrelevant.  
However, to account for the strong tendency of low-metallicity star-forming regions to produce many binary compact
objects, we adopt a power law
\begin{eqnarray}
\label{eq:LambdaVersusZModel}
\lambda(Z) &=& \lambda_o \text{min}[(Z/Z_\odot)^{-a}, F_{max}] 
\end{eqnarray}
with $a\in [0,3]$ and $F_{max}<10^3$
 (see, e.g.\cite{popsyn-LIGO-SFR-2008}, \abbrvKBLowZa).  For the purposes of illustration, we adopt a  concrete scale
factor  $\lambda_0 = 10^{-3}/M_\odot $,  a typical value suitable for neutron star compact binaries (see \abbrvPSgrb{},
\abbrvPSellipticals, and \abbrvKBLowZa).
To calibrate the exponent, based on Tables 2 and 3 of
\cite{popsyn-LowMetallicityImpact2-StarTrackRevised-2012} (model B), for binary black holes and black hole-neutron star
binaries, we adopt $a=1$, while for binary neutron stars we adopt $a=0$; see, e.g., their Table 1 and Figures
5-7. This choice of exponent provides an extremely conservative assessment of the impact of low Z
\cite[see,e.g.,][]{2012CQGra..29n5011O,gwastro-EventPopsynPaper-2016}. 
For binaries containing neutron stars, for simplicity and without loss of generality we adopt the universal delay time distribution
\begin{eqnarray}
\frac{dP_t(<t)}{dt} =  \begin{cases}
0 & t<10 \unit{Myr} \\
\frac{1}{t \ln (13 \unit{Gyr}/10\unit{Myr})} & t \in [10\unit{Myr},13\unit{Gyr}]
\end{cases} 
\end{eqnarray}
\abbrvPSgrb{} and \abbrvPSellipticals{} show this distribution is a reasonable approximation to compact binary delay
time distributions.\footnote{Simulations suggest the delay time distribution varies from model to model and with mass.
  These variations have less impact on our results than the evolving metallicity distribution of star forming gas.}
For binary black holes forming at metallicities $Z<0.25 Z_\odot$, we adopt the same prescription.  
For binary black holes formed near solar metallicity, the delay time distribution can favor long delays between birth
and merger, as demonstrated in Figures 9 and 10 of \abbrvPSellipticals{}.  [Figure 2 of  \cite{2016arXiv160508783L} is
  an extreme example of this well-known trend.] 
To be qualitatively consistent with detailed binary evolution calculations at near-solar metallicity (e.g., \abbrvPSellipticals{} and
\cite{gwastro-EventPopsynPaper-2016}), for 
black holes forming at metallicities $Z>0.25 Z_\odot$ we adopt a much more uniform delay time distribution, so 
coalescing black hole binaries have nearly uniform delay time distribution between $100\unit{Myr}$ and
$13\unit{Gyr}$; that said,  our conclusions are not sensitive to this choice.
For suitable choices of parameter, our phenomenological response function is qualitatively consistent with the results of detailed simulations of binary
evolution  \citep{2010ApJ...715L.138B,popsyn-LowMetallicityImpact2c-StarTrackRevised-2014,popsyn-LowMetallicityImpact2b-StarTrackRevised-2013,popsyn-LowMetallicityImpact2-StarTrackRevised-2012}.

Therefore, up to an irrelevant overall scale factor, the  present-day detection-weighted coalescence rate $r_D$  of binary compact objects formed by within two similar
galaxies  can be calculated via
\begin{eqnarray}\label{eqn:rate}
 r_D \propto  \int d\log Z  \int _{13 \unit{Gyr}}^0dt  V(Z) \lambda(Z) \frac{dP_t}{dt}(t) \dot{M}_* p(\log Z|t)
\end{eqnarray}

\section{Compact object binary formation rate}
\label{sec:results:BBH}

 Our four Milky Way-like galaxies have extremely similar star formation histories and metallicity evolution, particularly at late
 times.  
However, at early times, the \ExcitingGalaxy{} galaxy forms stars for $\simeq 2\unit{Gyr}$ at a lower characteristic metallicity
($Z\simeq 0.8 \times10^{-3}$) compared
to the \BoringGalaxy{} galaxy ($Z\simeq 2\times 10^{-3}$).  In this regime, the formation efficiency $\lambda$ and sensitive volume $V$
can depend sensitively on mass; for example, for binary black holes, the ratio of  $(\lambda V)_{\BoringGalaxy{}}/(\lambda
V)_{\ExcitingGalaxy{}} \simeq 3$.\footnote{Adopting a more extreme exponent for the metallicity dependence ($a=2$) only
  changes this ratio by of order $2$.}   However, in this same epoch, the star formation rate in the \ExcitingGalaxy{} (low-metallicity)
galaxy is smaller, by a factor of roughly 2.
Therefore, because only a fraction of order $10\%$  of all star formation occurs in this epoch, by this order of magnitude
argument, the overall number of
present-day coalescing binary black holes in our two galaxies is expected to differ by of order ten percent.
Using the concrete phenomenological calculations described above, we in fact find
$(r_{D}/M)_{\BoringGalaxy{}}/(r_{D}/M)_{\ExcitingGalaxy{}}\simeq 0.9$.   

The close agreement between the two galaxies' binary black hole populations, determined by the anticorrelation between
star formation rate and metallicity, may be a single example of a broad trend.  This anticorrelation could cause
galaxies with similar present-day properties to always have similar present-day binary black hole populations,
regardless of their detailed assembly histories.  If low metallicity star formation makes up a small fraction of the total stellar mass, which is the case for each Milky Way-like galaxy we study here, then the quantity $V(Z) \lambda(Z) \dot{M}_* p(\log Z|t)$ must differ by about an order of magnitude for the binary black hole merger rates to differ strongly.  Such a scenario is possible if we alter our conservative choice of the exponent $a$ in the $\lambda(Z)$ function, but nonetheless is rather unlikely that galaxies in this mass range will undergo drastically different early evolution. 
The effects described here
  are well within the scatter around the mass-metallicity relation.  Marginally different realizations of these
  histories can easily produce factors of order unity difference in galaxies with otherwise indistinguishable
  present-day properties.  

More broadly, Table \ref{tab:Results} shows the results of our calculations for the three types of compact binaries
described above.   These calculations show just how dramatically {\em different} the compact binary populations of galaxies
with different present-day {\em masses} 
could be.   The dwarf galaxies have an
exceptionally large fraction of low-metallicity star formation in their history \citep{Kirby13} relative to the more massive galaxies.  The precise details of their chemical
evolution can modify their present-day binary black hole binary populations by factors of order a few, despite  adopting
the conservative choices described above for the dependence of rate on metallicity.  The progenitors (and, if present,
electromagnetic counterparts) of future binary black hole gravitational wave events are overall much more likely to be located in nearby dwarf galaxies with a large population of low metallicity stars.

To highlight how sensitively compact binaries depend on detailed evolutionary trajectories, Table \ref{tab:Results}
shows results for binary neutron stars.  By our construction, no metallicity dependence is included in the present-day
event rate for neutron stars.  Thus, the differences in
present-day state between these two galaxy population models arise solely and exclusively on the time distribution history of past star formation.  In general, the present-day population depends often significantly (i.e., tens of
percent) on the assembly history alone, even aside from any composition-dependent effects.

\begin{table}
\begin{centering}
\begin{tabular}{llllc}\hline
\text{Simulation}  &  \text{BHBH} &  \text{BHNS}  & \text{NSNS} & M$_* (\msun)$\\
\hline
\hline
 \text{\BoringGalaxy} & 0.0224 & 0.000247 & 0.000206  & $4.24 \times 10^{10}$ \\
 \text{\ExcitingGalaxy}  &0.0216 & 0.000297 & 0.000228 & $4.46 \times 10^{10}$ \\
  \text{h239}  &  0.023 & 0.000354 & 0.000268 & $4.50 \times 10^{10}$\\
 \text{h285} &  0.0236 & 0.000284 & 0.000241 & $4.56 \times 10^{10}$ \\ \hline
 \text{\DwarfOne} &  0.0381 & 0.000308 & 0.000441& $7.8 \times 10^9$ \\
 \text{\DwarfTwo}  & 0.0949 & 0.000257 & 0.000884& $2.5 \times 10^8$ \\
\hline
\end{tabular}
\end{centering}
\caption{\label{tab:Results}Event rates per unit mass $r_D/M$, arbitrary units, for the three different types of binaries discussed, along with the stellar mass of each galaxy.  
}
\end{table}

\section{Implications for transient multimessenger astronomy}
\label{sec:Discussion}

While a gravitational wave detection provides detailed information about merging objects (i.e. masses, spins, distance), we need further knowledge to understand the actual origins of the progenitors.  Even when electromagnetic counterparts are available, the long delay times between binary formation and merger limit the prospects of examining the environment where the binary first formed.  On the one hand, compact binaries are kicked by supernovae, moving substantially away from their birth position
\citep{2013ApJ...776...18F,2014ARAA..52...43B}. 
On the other hand, particularly during the early epoch of galaxy formation, gas in galaxies is well-mixed: stars and adjacent gas generally do not have similar chemical composition.  These mixing effects have been previously recognized as an obstacle to interpreting transient event spectra.  For
example,   \citet{2010MNRAS.402.1523P} previously demonstrated that absorbing gas neighbouring transient events (there, long GRBs)
would generally have high metallicity, even for low-metallicity progenitors.   
\citet{2010MNRAS.402.1523P} have previously used hydrodynamical simulations to demonstrate that observed ambient
metallicities (there, using damped Lyman $\alpha$ absorbers in the host) do not tightly constrain the metallicity
distribution of the progenitor; see, e.g., their Fig 3.  Thus, the metallicity of stars and gas adjacent to a specific merger event provides
few direct, unambiguous clues to a compact binary merger's progenitors.

Fortunately,  with the advent of
IFUs and position-resolved spectroscopy, observers can now probe the star formation history and metallicity of individual gas packets at different points in a galaxy.  
These highly-detailed probes will be essential tools to develop a comprehensive model of the galaxy's merger and
chemical evolution history.    Obtaining the galaxy-wide evolutionary history can help us infer compact binary formation
conditions by identifying the lowest-metallicity formation events using stellar archaeological techniques.  
These techniques have been applied with great success to other transient events. 
For example, in several cases the metallicity of gas neighbouring a long GRB (\citet{ 2008AJ....135.1136M};
\citet{2010AJ....140.1557L}) has been directly measured.  
The precise host offset can be compared to the distribution of light and star formation  \citep{2010ApJ...708....9F}.
Finally,  on a host-by-host basis, the delay time between birth and merger has been constrained for
short GRBs \citep{2010ApJ...725.1202L}
 and
  SN Ia \citep{2011MNRAS.412.1508M}; see, e.g., the review in  \cite{2014ARAA..52...43B}.

Thus, despite the challenges involved in obtaining sufficient statistics, possibly requiring third-generation instruments to
obtain sufficiently many associations, past experience suggests  host galaxy associations
will provide unique clues into the formation mechanism of compact binaries.
Each host galaxy associates a merger to a unique star formation history and metallicity distribution.
With many events, these associations can potentially determine the ``response function'' for compact binaries: how often star forming gas of a given metallicity evolves into
merging compact binaries.

\section{Conclusions}
\label{sec:conclude}

We examine the present-day populations of coalescing compact binaries in galaxies with different assembly histories.  We
combine  detailed and state-of-the-art cosmological simulations of galaxies with a simple but robust phenomenological
model for how compact binaries form from different environments.   
 We demonstrate that
galaxies which appear similar at $z = 0$ but have differing merger histories will have somewhat different
detection-weighted compact binary
coalescence rates.

For binary black hole mergers in particular, we show that the present-day binary black hole coalescence rate for our two
Milky-Way like galaxies is nearly identical, independent of their highly distinctive early-time formation histories.   
This result perhaps comes as somewhat of a surprise, considering the early differences in stellar and chemical evolution.
Our calculations adopt the same framework as  prior investigations, which demonstrated 
that black hole merger rates depend sensitively on low-metallicitiy environments
\citep[see,e.g.][]{PSellipticals,popsyn-LowMetallicityImpact2b-StarTrackRevised-2013,gwastro-EventPopsynPaper-2016}.   More broadly,
because compact binaries can merge long after they form, their host galaxy can evolve substantially in composition
between birth and merger.
Nonetheless, the apparent anticorrelation between the star formation rate and metallicity evolution of our galaxies has
led to similar late-time populations, despite substantial differences early on.
Further investigation is critical to assess whether this similarity is retained for more generic galaxy assembly
histories and binary formation models.    Now that we have introduced this method as a proof-of-concept here, it can be applied to large volume simulations, such as Illustris, EAGLE, or Romulus \citep{Illustris,EAGLE,Tremmel16}, in order to examine large numbers of galaxies and thus obtain statistically significant results.

The detailed analysis of the compact binary populations formed through the assembly history of individual galaxies is
complementary to the  population-based approach reported in
\cite{2016arXiv160508783L}.     More broadly, our analysis reflects  similar theoretical studies performed in the interpretation of, for
example, long GRBs and their host galaxies. 
For example, \cite{2009ApJ...702..377K} demonstrated that a sufficiently strong bias towards low-metallicity star formation would predict
most  events in the local universe occur low-mass and dwarf galaxies.
For less extreme metallicity biases, subsequent calculations by  \citet{2011MNRAS.417..567N} demonstrated that the
metallicity distribution within galaxies will usually lead to events in a wide range of host galaxies in the local universe.

In our investigation, we have neglected the impact of population III stars.
Previous studies suggest the first generation of stars may contribute a small but nonzero fraction of detectable events,
potentially with distinctively high black hole masses
\citep{2016MNRAS.460L..74H,2016PhRvL.117f1101S,2014MNRAS.442.2963K}.

In short, in this work we have demonstrated that  the confounding effects of host galaxy assembly history can in the
near term complicate the interpretation of associations between GW sources and candidate host galaxies.   However, given sufficient
statistics, as will inevitably become available with next-generation GW instruments, combined with large-scale
multiband followup, these confounding challenges can  both be overcome and converged into opportunity.   In the far future,  with hundreds of thousands of thousands to millions of events per year in
networks like Cosmic Voyager, Einstein Telescope, and DECIGO, GW measurements could even  provide complementary statistical probes of the past history of
galaxy assembly and evolution.

\section*{Acknowledgements}
ROS acknowledges support from NSF award AST-1412449, via subcontract from the University of Wisconsin-Milwaukee, and PHY-1505629.
JB acknowledges generous support  from the Helen Gurley Brown Trust.
A portion of this work was performed at the Aspen Center for Physics, which is supported by National Science Foundation grant PHY-1066293.


\end{document}